# Stability of rhombohedral phases in vanadium at high-pressure and high-temperature: first-principles investigations

Yi X. Wang,[1,2] Q. Wu,[1] Xiang R. Chen,[2,*] and Hua Y. Geng[1,*]

**Abstract:** The pressure-induced transition of vanadium from BCC to rhombohedral structures is unique and intriguing among transition metals. In this work, the stability of these phases is revisited by using density functional theory. At finite temperatures, a novel transition of rhombohedral phases back to BCC phase induced by thermal electrons is discovered. This reentrant transition is found not driven by phonons, instead it is the electronic entropy that stabilizes the latter phase, which is totally out of expectation. Parallel to this transition, we find a peculiar and strong increase of the shear modulus $C_{44}$ with increasing temperature. It is counter-intuitive in the sense that it suggests an unusual harding mechanism of vanadium by temperature. With these stability analyses, the high-pressure and finite-temperature phase diagram of vanadium is proposed. Furthermore, the dependence of the stability of RH phases on the Fermi energy and chemical environment is investigated. The results demonstrate that the position of the Fermi level has a significant impact on the phase stability, and follows the band-filling argument. Besides the Fermi surface nesting, we find that the localization/delocalization of the *d* orbitals also contributes to the instability of rhombohedral distortions in vanadium.

[1]National Key Laboratory of Shock Wave and Detonation Physics, Institute of Fluid Physics, CAEP; P.O. Box 919-102, Mianyang 621900, Sichuan, People's Republic of China. [2]College of Physical Science and Technology, Sichuan University, Chengdu 610064, China. *Correspondence and requests for materials should be addressed to H.-Y.G. (email: s102genghy@caep.ac.cn) or X.-R.C. (email: xrchen@scu.edu.cn).

**Introduction**

Revealing and elucidating the trend of structural transformations and the underlying mechanisms in elemental metals is a fundamental topic in condensed matter physics. In recent years, the transition metal vanadium has attracted much experimental[1-3] and theoretical[4-7] attention because of its important applications and the puzzling softening in the strength and the subsequent transition from the body-centered cubic (BCC) phase to the low-symmetry rhombohedral (RH) structures driven by pressure.

The first direct indication of a phase transition in vanadium perhaps came from the theoretical observation of a strong softening in the transverse acoustic phonon mode along the **Γ-H** direction at high pressures, which eventually becomes imaginary when beyond 130 GPa.[8] This softening is corroborated by the calculated elastic constants where the shear modulus $C_{44}$ continuously decreases to zero and becomes negative, suggesting instability of the BCC structure.[9,10] This bizarre behavior was believed due to Fermi surface nesting and the Kohn anomalies, but the resultant crystalline structure was not proposed at that time. In fact, there had an early experimental study of vanadium up to 154 GPa, but no phase transition was reported.[11] Later, Ding et al.[12] conducted delicate x-ray diffraction experiments using diamond anvil cell (DAC) up to 150 GPa, and found a novel RH phase, which is a slight distortion of BCC structure and appears from about 63-69 GPa. It was soon confirmed by Lee et al. using static lattice density functional theory (DFT) calculations, and showed that the RH phase was the ground state when beyond 84 GPa.[13,14] They also predicted other two phase transformations that were not detected in Ding's experiment, i.e., a transformation to another RH structure at 120 GPa and the third transformation back to the BCC structure at 280 GPa. Other theoretical studies also reported qualitatively similar findings.[15,16] Lattice dynamics calculations also supported the argument that the RH phase should transform back to BCC structure when above 250 GPa.[17] On the other hand, Jenei et al.[18] reported a BCC→RH transition at 30 GPa when no pressure medium was used, whereas it was at about 60 GPa if in the Ne pressure medium. It seems non-hydrostatic condition has a strong impact on the transition pressure. Nonetheless, in Ding et al.'s experiments, the transition pressure $P_c$ is at 69 GPa if without a pressure medium, and $P_c$ = 63 GPa if in He pressure medium.[12] Therefore it is possible that the deviation in $P_c$ might not be due to the non-hydrostatic condition, other factors such as chemical environment might also have some contributions. In addition, the energetics stability of the RH phase of vanadium with respect to high symmetry candidate structures, such as FCC, HCP, simple cubic, and simple hexagonal structures has been reported in literature,[15,19,20] and the RH phases are the most stable structures within our studied pressure range. For this reason, we will not discuss these structures below.

It has established a theoretical consensus on the phase transition sequence of BCC→$RH_1$→$RH_2$→BCC in vanadium under pressure. But there still are discrepancies on the exact values of the transition pressure. One of the most remarkable difference was reported by Qiu *et al*.,[16] in which the first transition pressure was evaluated to be 32 GPa, much smaller than all other estimations. Qiu *et al*.'s calculations were under hydrostatic condition, thus are irrelevant to Jenei *et al*.'s argument.[18] Except the value reported by Qiu *et al*., all other theoretical transition pressures of BCC→$RH_1$ were located between 60-84 GPa. By contrast, the predicted $RH_1$→$RH_2$ transition pressures scattered between 115-160 GPa, and the predicted $P_c$ of $RH_2$→BCC was between 240-297 GPa. The reason underlying this large uncertainty is unknown. To clarify this theoretical deviation is important for understanding the discrepancy in experimental transition pressures. In this work, we will revisit the phase transition and structural stability in vanadium with highly accurate *ab initio* calculations, in attempt to shed new light on this open issue.

On the other hand, most investigations on this problem reported in literature mainly focused on the ground state. Few study was devoted to finite temperatures. Recently, Landa *et al*.[21] studied the stability of vanadium metal and vanadium-chromium alloys in BCC phase at high temperatures and pressures by calculating the anharmonic phonon dispersions. Their results showed that lattice vibrations slightly weaken the stability of RH phases; but $RH_1$ is stable up to the static melting curve, and $RH_2$ is stable beyond the melting point estimated by shock wave experiments (for vanadium there is a large discrepancy between the melting curves determined by static compression techniques such as DAC[22] and dynamic shock wave method[23]). Nevertheless, as revealed in previous investigations, the anomalous softening and phase transition in vanadium originate in electronic structure, and closely relate to the Fermi surface nesting, hence the contribution by lattice dynamics might be secondary. By intuition, thermo-electrons should have greater contributions, which to our best knowledge have not been studied yet. Whether the electronic temperature has significant influence on the shear modulus and phase stability of vanadium is still unknown. For this reason, a thorough and comprehensive investigation on thermo-electron effects will also be performed in this work. Furthermore, we will elucidate the possible impact of the Fermi level position and charge transfer on the softening behavior and phase stability of vanadium. This will shed light on how the compression behavior of vanadium can be changed by surrounding chemical environment, which might be helpful in understanding the large discrepancy in experimental transition pressures.

**Results and Discussion**

**Phase transition at zero Kelvin.** Since Jenei *et al.* observed a low transition pressure of $P_c$ ~30 GPa under non-hydrostatic condition,[18] and Qiu *et al.* also reported a similar hydrostatic $P_c$ with theoretical calculations,[16] it becomes necessary to revisit the BCC→RH→BCC transition problem with careful treatment. Qiu *et al.* attributed their discrepancy with respect to other estimates in $P_c$ to the constraints imposed by the hypothetical transition pathway that was employed in those works.[13,14,24] Jenei *et al.* also speculated that the thermodynamic equilibrium $P_c$ of the first transition might be at ~30 GPa, and concluded that Ding *et al.*'s value of ~60 GPa could be due to a kinetic effect where a large energy barrier hinders the transition. In order to figure out the possible reason of such a big scattering in theoretical calculations, a comparative investigation will be performed in this work, including both the methods to locate the transition pressure and the accuracy of DFT calculations.

We first optimize the structures at zero Kelvin with all parameters being fully relaxed. We choose $RH_1$ and $RH_2$ as the initial structures, and optimize them at different pressures without any symmetry constraint being imposed. The calculated enthalpy difference with respect to the BCC phase, together with the variation of the structural parameter (angle $α$) in $RH_1$ and $RH_2$ structures as a function of pressure, are shown in Fig. 1. These fully relaxed results clearly reveal that the equilibrium transition of BCC→$RH_1$ is not at 30 GPa, where $RH_1$ is dynamically instable and spontaneously collapses to BCC phase. As shown in the figure, the metastable region of $RH_1$ extends down to 50 GPa. At lower pressures it becomes highly unstable. Nonetheless, its relaxation back to the BCC phase is not perfect. The residual angle $α$ is about 109.51° at 20-40 GPa. This reflects a fact that the potential well of BCC is flat, and non-hydrostatic loading can easily drive vanadium towards RH-like deformations. It is necessary to point out that the angle obtained in Jenei *et al.*'s experiment is $α = 109.61°$, which is far less than Lee *et al.*'s theoretical value of $α = 110.25°$, but close to the imperfect BCC that collapsed from $RH_1$ as shown in Fig. 1. Also note that the $RH_2$ phase leads to a similar distorted BCC structure with $α = 109.39°$ when below 110 GPa. In our calculations, the $RH_1$ phase becomes the ground state when beyond 98 GPa. It has a bigger angle $α = 110.17°$, compared to $α = 109.47°$ of the perfect BCC structure. Transition to another RH phase ($RH_2$) occurs at ~128 GPa, which attains the maximum stability at ~210 GPa, with $α = 108.23°$. $RH_2$ is the ground state up to ~284 GPa. As the pressure increases further, it becomes metastable and eventually collapses back to the BCC structure at ~300 GPa. On the other hand, the metastability of $RH_1$ phase extends up to ~247 GPa, where it also automatically transforms back to BCC. Our results are in good agreement with Ref. [13], except a small numeric difference on the phase boundaries. This indicates that Lee *et al.*'s method is compatible with the full structural relaxation.

We also study the phase transition using the same method as Lee *et al.*[13] (i.e., the method II in Sec. Methods). This comparative calculation provides a consistent benchmark for the transition pressures. In this method, the unit cell volume is conserved. Qiu *et al.* argued that this treatment would give rise to a higher transition pressure in Lee *et al.* and others' calculations.[16] Nonetheless, our results with the full relaxation discussed above suggested that this argument might not be true. The error introduced by fixing the volume at $V_0$ can be corrected using the formula

$$H(\delta, P_0) \approx U(\delta, V_0) + PV_0 - \frac{1}{2B(\delta, V_0)} \Delta P(\delta, V_0)^2 V \quad (1)$$

It corrects the enthalpy along the deformation path. In Eq. (1), $\Delta P(\delta, V_0)$ is the change of the pressure at a given volume $V_0$ and a rhombohedral deformation magnitude $\delta$, $B$ is the bulk modulus. In Lee *et al.*'s evaluation, only the first term was used.[13] We carefully tested and found that the correction from the third term is indeed very small. This supports Lee *et al.*'s assessment that the internal energy $U(\delta, V_0)$ is sufficient when studying the relative phase stability, and the contribution from volume relaxation can be safely ignored.

Our enthalpy differences as a function of the rhombohedral distortion $\delta$ at several pressures calculated with method II are plotted in Fig. 2. It clearly shows that the pressure-induced transformations of BCC→RH$_1$, RH$_1$→RH$_2$, and RH$_2$→BCC occur in sequence with increasing pressure. A small energy barrier between equilibrium transitions is evident. Therefore the transitions should be first-order rather than continuous. In this method, the BCC is the only stable structure at low pressures. It changes to RH$_1$ at about 103 GPa. The stability of RH$_2$ structure gets enhanced with increased pressure, and becomes the ground state at ~126 GPa. The RH$_2$ structure attains the maximum stability at about 211 GPa, beyond which the enthalpy well becomes shallow. When above 300 GPa, the RH$_2$ phase disappears and the BCC becomes the only stable structure again. It should be noted that these results are in good agreement with those obtained by the method I. Also Fig. 2 clearly illustrates that there is no local enthalpy minimum when below 60 GPa, and thus no metastable region for RH$_1$ phase. This observation disproves Jenei *et al.*'s conjecture that the equilibrium $P_c$ of BCC→RH$_1$ is at ~30 GPa.

In order to acquire a comprehensive understanding of these results, we summarize the critical pressures in Table I. For comparison the theoretical results by Lee *et al.*,[13] Qiu *et al.*,[16] and Verma *et al.*[15] are also listed. It is evident that both our results calculated with method I and method II are consistent very well with each other. For this reason, Qiu *et al.*'s comment on Lee *et al.*'s results is not pertinent.[16]

Compared with Lee *et al.*, our $P_c$ of BCC→RH$_1$, RH$_1$→RH$_2$, and RH$_2$→BCC are slightly higher by about 14, 9, and 4 GPa, respectively. Verma *et al.*[15] gave a BCC→RH$_1$ transition pressure of 60 GPa, which is close to Ding *et al.*'s experimental value, but their transition pressures of RH$_1$→RH$_2$ and RH$_2$→BCC are far away from others' estimate. Landa *et al.*[24] predicted a similar BCC→RH$_1$ transition pressure, but no other two transition pressures were reported, therefore we do not discuss their results here.

In order to understand the discrepancy in these theoretical data, we notice that for the latter two transitions, the deviation in $P_c$ among Qiu *et al.*, Lee *et al.*, and our results is small. It is just 11 and 19 GPa for RH$_1$→RH$_2$ and RH$_2$→BCC transition, respectively. In contrast, the deviation in BCC→RH$_1$ is as big as 71 GPa. By inspecting Figs. 1 and 2, we find that the enthalpy difference of RH$_1$ is much smaller than RH$_2$ phase. The small enthalpy difference requires very accurate calculation in order to obtain a reliable estimate of the transition pressure. For example, though our $P_c$ of BCC→RH$_1$ is about 103 GPa and larger than Lee *et al.*'s 84 GPa, the enthalpy difference between RH$_1$ and BCC at 84 GPa is less than 0.37 meV per atom. This is almost the accuracy limit of DFT based on pseudopotential method. We notice that both ours and Lee *et al.*'s calculation already achieved the absolute computation convergency, but we used a different PAW pseudopotential newly released with VASP. This could be the reason for the minor numeric difference. Except this, our results are in good agreement with theirs.

The shallow depth of the enthalpy well in RH$_1$ implies that its transition pressure is sensitive to the computation accuracy. In our calculations, we find that a $P_c$ of 84 GPa for BCC→RH$_1$ could be obtained if reduced the *k*-point mesh to $30\times30\times30$. A lower $P_c$ would be obtained if further degrade the computational accuracy. By comparison, Verma *et al.*[15] used a coarse *k*-point grid of $18\times18\times18$, and obtained a low $P_c$ for BCC→RH$_1$ (60 GPa) and a high $P_c$ for RH$_1$→RH$_2$ (160 GPa) transition. Qiu *et al.*[16] employed FLAPW method as implemented in WIEN2k_07 package. They predicted a BCC→RH$_1$ transition pressure of ~32 GPa, which is much smaller than Ding *et al.*'s experimental value[12] and all other theoretical studies.[13-15] As noted above, they ascribed this discrepancy to the constraints imposed by the transition pathway employed in calculations. But our above analysis showed that this is not the case. The true reason might be the computation precision that was not clearly specified in Ref. 16.

By concluding this subsection, we emphasize that the scattering in the theoretical data of $P_c$ mainly comes from computational accuracy. We found that there is no local enthalpy minimum and metastable region for $RH_1$ phase at around 30 GPa if fully convergent DFT calculations were performed. The non-hydrostatic effects reported by Jenei *et al*. might be due to unexpected exploration of angle $\alpha$ away from the ideal value of 109.47° by shear deformations, rather than transformation into the true $RH_1$ phase. We thus suggest to reinvestigate this problem experimentally by releasing pressure from the stable $RH_1$ phase with $\alpha = 110.17°$. This would reveal the distinct collapse of $RH_1$ back to BCC structure and give a lower bound of its metastability.

**Thermo-electron effect.** The proposed Fermi surface nesting mechanism for the instability of BCC structure of vanadium[10] implies that the transformation between BCC and RH phases should sensitively depend on subtle features of the Fermi surface. Usually there are three factors that will affect the Fermi surface structure: (i) lattice dynamics might modify the Fermi surface via electron-phonon interaction; (ii) thermo-electron excitations will blur and smear the subtle features in the Fermi surface; (iii) shift the Fermi level will also change the features of the Fermi surface. It is known that increasing the temperature leads to excitation of lattice vibrations and thermo-electrons. Recently, Landa *et al.*[21] studied phonon effects on the phase stability of vanadium metal and vanadium-chromium alloys at high temperatures and pressures. Their results indicated that phonon has little impact on the stability of RH phases. Based on their conclusion, here we will mainly discuss the thermo-electron effects by using the finite temperature DFT method.[25] The effects of Fermi level position will be discussed in next subsection.

Different from lattice dynamics, we find that thermo-electrons have a very strong impact on the stability of RH phases. Our calculated results for various temperatures at a pressure of 211 GPa are plotted in Fig. 3, at this pressure the RH$_2$ phase attains its greatest stability. It is evident from this figure that the electronic temperature strikingly weakens the RH stability. The RH$_2$ phase transforms back to BCC structure when the temperature is above ~1915 K at 211 GPa. In comparison, it would be stable up to 8000 K at 182 GPa if including only the phonon contribution, as Landa *et al*. reported.[21] The thermo-electron effect is much more important than phonons. A simple analysis shows that including both thermo-electronic and phonon corrections will further reduce the transition temperature by ~260 K at 200 GPa (see Supplementary information). This phenomenon is compatible with the Fermi surface nesting mechanism of the BCC→RH transition, in which the temperature smears the Fermi surface and thus stabilizes the BCC phase. Nonetheless, as will be shown below, the Fermi surface nesting might not be the only mechanism for the BCC stabilization at high temperatures.

In order to deepen our understanding about this temperature-induced reentrant transition, we compute the temperature dependent electronic density of states (DOS) of RH$_2$ phase at 211 GPa, and show them in Fig. 4. The DOS of BCC phase at 580 K is also plotted for comparison. We can see that the BCC phase has more localized electrons (having *d* orbital character, at just below the Fermi level), whereas in RH phases the electrons are more delocalized. As the temperature increases, the valence electrons delocalize further and give a smooth DOS. This is in line with the expectation. The Fermi-Dirac distribution function $f(\varepsilon,T)$ is also plotted as an inset in Fig. 4, from which we can see that only a narrow region near the Fermi level (±0.5 eV) is directly affected by quantum statistics. However, its indirect influence on the whole DOS is remarkable and extends down to lower energy, mainly due to the electron-electron correlation that is automatically included in the self-consistent field procedure of the finite temperature DFT. For example, a noticeable modification on DOS still can be observed at -4 eV (measured from the Fermi level). The indication is that the classic argument based on perturbation theory, i.e., the temperature effects on electronic properties is usually restricted to an energy scale of $k_B T$ around the Fermi level,[26] is invalid. There are strong electron-electron correlations involved in vanadium. From this perspective, the stability of RH phases with respect to BCC might not solely be due to the Fermi surface nesting. Note that both Fermi surface nesting and band Jahn-Teller effect will result in a structure with a low DOS at the Fermi level. But at a temperature of 2321 K where BCC phase is favored, our calculations show that the DOS of the BCC phase at the Fermi level is higher than that of RH phases. This is at odds with the proposed Fermi surface nesting mechanism. Furthermore, with increasing temperature, the DOS of RH phases at the Fermi level decreases, but its stability becomes weakened instead, which is also incompatible with the picture of Fermi surface nesting or the Jahn-Teller mechanism. As shown in Fig. 4, the thermo-electrons effects extend down to a low energy region, therefore it is natural to suspect that the transition back to BCC at high temperatures is caused by the states on the Fermi surface.

To elucidate that, we calculate the band energy and thermo-electronic entropy difference between BCC and RH phases at elevated temperatures. The results for $RH_2$ at a pressure of 211 GPa are given in Fig. 5. Unexpectedly, we find that the band energy contribution is almost independent of the temperature, even though the modification on the DOS by temperature is remarkable. Instead, it is the *electronic entropy* that shifts the free energy difference up and finally stabilizes the BCC structure. Therefore we can conclude that the BCC phase at high temperatures is favored by electronic entropy, rather than by changes to the Fermi surface. A subsequent question is why entropy prefers BCC instead of RH? The answer is at Eq. (11) (in Sec. Methods), from which one can see that the entropy maximizes if a large DOS presents at or near the Fermi level. In other words, the more the electronic orbitals localize to the Fermi level, the greater the electronic entropy becomes. As depicted in Fig. 4, BCC structure has more localized *d* states and higher DOS near the Fermi level, hence it has larger electronic entropy and finally becomes the most thermodynamically stable phase at high temperatures. This mechanism is completely different from the band Jahn-Teller effect, Fermi surface nesting, or electronic topological transition that were proposed for the pressure-induced BCC→RH transition.

**Phase diagram and elastic constants.** Based on the calculated stability of BCC, $RH_1$, and $RH_2$ phases at high pressure and temperature, we construct a phase diagram for vanadium. This is the first time that such a comprehensive diagram of this metal is proposed. As shown in Fig. 6, BCC is the only stable solid phase up to 98 GPa (estimated with our BCC→$RH_1$ transition pressure) and stands up to the melting point. With increasing pressure and temperature, $RH_1$ phase becomes stable when below 1440 K and 140 GPa. The stable region of $RH_2$ phase locates between 126-280 GPa, with the maximal transition temperature of 1915 K at 211 GPa. As the temperature increases, both $RH_1$ and $RH_2$ phases become unstable and transform back to BCC, which then melts to the liquid phase. This picture drastically changes our previous understanding that $RH_1$ could stand up to the static melting curve[22] and $RH_2$ is stable up to the shock wave melting curve.[23] It is interesting to note that at the low pressure side, the phase boundary slope of BCC→$RH_1$ and $RH_1$→$RH_2$ is positive. This indicates that $RH_1$ is denser than BCC phase and $RH_2$ is denser than $RH_1$ phase, according to the Clausius-Clapeyron equation. When the pressure is higher than 211 GPa, the slope of the $RH_2$→BCC phase boundary becomes negative, suggesting that under this thermodynamic condition BCC becomes denser than $RH_2$ phase. The BCC-$RH_1$-$RH_2$ triple point is determined at 1440 K and 140 GPa, where spectacular mechanic properties could be expected. The phase diagram as shown in Fig. 6 is unique. As discussed above, it is a consequence of competition between two different transition mechanisms: along the compression direction Fermi surface nesting destabilizes BCC phase, whereas along the temperature dimension, the BCC phase is favored by electronic entropy.

The competition of the Fermi surface nesting and electronic entropy implies that the strength of vanadium should increase with increasing temperature. This is an astonishing prediction and at odds with widely accepted (though empirical) experience that temperature always *softens* metals. To verity this, we also calculate the elastic constants of vanadium at elevated temperatures and pressures.

The calculated elastic moduli $C_{44}$ and $C'$ as a function of pressure at zero Kelvin are shown in Fig. 7. For comparison the results of Landa *et al*.[10] and the experimental data at ambient conditions[26] are also included. It is evident that our results are in good agreement with the experimental data and Landa *et al.*'s full-potential linear muffin-tin orbitals (FPLMTO) results. We find $C_{44}$ becomes negative between 125 and 260 GPa, at the same time $C'$ is anomalously softened within the same pressure range (similar results were observed by using the exact muffin-tin orbitals (EMTO) calculations[10]). This is in sharp contrast to the theoretical $C_{44}$ values reported by Qiu *et al.*[16] (not shown here), which are far from ours and Landa *et al.*'s results. Qiu *et al.* argued that the discrepancy is due to the pressure correction neglected in Landa *et al.*'s work. However, here we used the same formula as Qiu *et al.* and included the pressure correction explicitly when calculating the elastic moduli $C_{44}(p)$ and $C'(p)$. The perfect match between ours and Landa *et al.*'s results unequivocally demonstrates that the pressure correction is not the main reason of their discrepancy. Rather, as already mentioned in previous subsection, the computation convergence quality might be the true cause.

The electronic temperature effects on the shear modulus $C_{44}$ of BCC phase as a function of pressure at different temperatures are plotted in Fig. 8. As we envisioned above, in the whole pressure range we studied, $C_{44}$ indeed increases with the temperature, and stabilizes the BCC against RH phases. This phenomenon originates in electronic structures and is against our empirical intuition. It should be noted that when temperature is above ~1900 K, the mechanically instable region (with negative $C_{44}$) of BCC phase disappears completely. This is consistent with the phase diagram as drawn in Fig. 6. Therefore we discover a novel temperature-induced *hardening* phenomenon in vanadium, which is very rare (if any) to our knowledge. At ambient pressure, when temperature increases from room temperature to 2000 K, $C_{44}$ increases by 13.18%. At 50 GPa and 300 GPa where BCC is always stable, when temperature is increased from room temperature to 3000 K, $C_{44}$ increases by 75.43% and 53.21%, respectively, as shown in the inset of Fig. 8. At high enough temperatures, especially when near the melting point, the thermal motion of nuclei will soften the metal via thermal activation mechanism. Therefore we predict that the strength and shear modulus of vanadium will increase to a maximum, and then drop down to zero with increasing temperature.

**Effect of Fermi level and charge transfer.** This subsection is devoted to the charge transfer induced by chemical environment and the subsequent shift of the Fermi level. Both Fermi surface nesting and band Jahn-Teller mechanism depends sensitively on the position of the Fermi energy or the electronic chemical potential $E_F$. By using the partial jellium model briefed in Sec. Methods, we adjust the position of $E_F$ slightly and investigate its impact on phase stability of vanadium. In Fig. 9, the enthalpy difference as a function of the rhombohedral deformation parameter $\delta$ are shown for three shifted Fermi levels at a pressure of 211 GPa. Shifting the Fermi level means to change the orbital occupations at the valence band top or conduction band bottom by charge transfer. At 211 GPa, the RH$_2$ phase attains the maximum stability. If the RH structures are stabilized by $s \rightarrow d$ electronic transition or band Jahn-Teller effect, it should have the optimum orbital occupation at this pressure. Therefore, lifting (adding electrons) or descending (removing electrons) the Fermi level moves the system away from the optimum occupation, thus both should weaken the stability of RH phases. Nonetheless, our calculations show that at 211 GPa shifting down the Fermi level stabilizes the RH phases further, whereas shifting it up destabilizes RH greatly. This is in line with Landa *et al.*'s band-filling argument when alloying vanadium with the same transition series,[24] but is incompatible with Jahn-Teller mechanism. However, this observation is not contrary to the Fermi surface nesting, which should depend sensitively on the subtle Fermi surface structure, and thus its position.

Furthermore, our calculation indicates that the maximal stability of RH$_2$ phase at this pressure is attained when $\Delta = -2.15\%$. Here $\Delta$ is defined as the total charge percentage being added/removed to the system. Further shifting down the Fermi level destabilizes the RH phases. When $\Delta < -4.85\%$, the BCC phase becomes favorable again. On the other hand, BCC also becomes stable when $\Delta > 0.77\%$. The large stable range of $\Delta$ for RH$_2$ phase implies that the Fermi surface nesting alone can not be the distortion mechanism. In the inset of Fig. 9, we plot the calculated differential charge density between $\Delta = -0.77\%$ and $\Delta = 0$. It is evident that the removed electrons (or added holes) are well localized around the nuclei. Analysis indicates that they are mainly *d* character. The variation of enthalpy difference with Δ shows that within this pressure range the RH$_2$ phase of vanadium dislikes electrons, and has a low electronegativity with respect to the BCC phase. Such behavior should be a consequence of *d* orbitals delocalization. Therefore localization/delocalization of the *d* electrons also has an important role in RH stability even at low temperatures.

In Fig. 10, the variation of the electronic DOS of RH$_2$ phase for three shifted Fermi levels at 211GPa are displayed. It can be seen that the modification on DOS by Fermi level shifting is very small. Based on Figs. 9 and 10, it seems that the RH$_2$ phase becomes more stable when the DOS at Fermi level is higher. This is at odds with the band Jahn-Teller mechanism and the Fermi surface nesting. However, when compared to the BCC phase, the relative DOS difference at the Fermi level for Δ = -0.77%, Δ = 0, and Δ = 0.77% are 0.34, 0.22, and 0.05, respectively. In other words, when the Fermi level shifts up, the relative DOS difference at the Fermi level along the rhombohedral deformation path decreases. At the same time, the stability of RH$_2$ phase continuously weakens, and eventually transforms back to the BCC phase.

Consider now how the band structure responses when the Fermi level being shifted. In Fig. 11, the band structures of vanadium in RH$_2$ and BCC phases at 211 GPa are plotted for three shifted Fermi levels. It can be seen that the band structure is just slightly modified when the Fermi level being changed, and is consistent with the DOS shown in Fig. 10. When comparing the band structure of RH$_2$ to BCC phase, we find that the rhombohedral distortion splits the levels with $t_{2g}$ symmetry at **Γ** and **H** points, which is consistent with the results reported by Landa *et al.*[10] and Ohta *et al.*[27] The interesting finding here is that though all Fermi levels are well positioned within the pseudo-gap opened by splitting of the $t_{2g}$ states at **Γ** point, the RH$_2$ stability is changed differently. This intriguing phenomenon suggests that other factors besides the splitting of the $t_{2g}$ might be involved in BCC→RH phase transformation.

From above analyses, it is clear that the Fermi level position, as well as accompanying charge transfer, has a strong effect on the phase stability of vanadium. Since electron topological transition has been excluded as the main driving force for the BCC→RH phase transitions,[28] our calculations suggest that the Fermi surface nesting, the distortion induced band splitting, and localization/delocalization of the *d* orbitals might be responsible for these unique phase transitions in vanadium, as well as the unique mechanical properties.

**Conclusions**

In summary, a thorough and comprehensive theoretical study of the phase transition in vanadium at high pressure and high temperature has been carried out with first-principles calculations based on density functional theory. It was found that the scattering in the theoretical pressure of the first-order phase transitions (BCC→$RH_1$, $RH_1$→$RH_2$ and $RH_2$→BCC) is mainly due to computation convergence quality. With high enough precision computations, the transition pressures were pinned to 98, 128, and 284 GPa, respectively. Our calculations also predicted no local minimum and metastable region for $RH_1$ at low pressures, and suggested that the transition pressure of 30 GPa reported by Jenei *et al*. might not be due to BCC→$RH_1$ transition, rather it could arise from shear deformations. Considering the complication arisen from non-hydrostatic effects, we suggest to investigate the stability of $RH_1$ along the pressure releasing path, by which the abrupt collapse of $RH_1$ to BCC at ~60 GPa might be more distinct to observe.

The thermo-electron effect on the stability of vanadium in different structures was studied by using the finite temperature DFT. We observed that both phonon (see Ref. 19) and thermo-electronic effects reduce the stability of RH phases, but the effects of thermal electrons are more important. Both $RH_1$ and $RH_2$ phases transform back to BCC structure when the temperature is above 1440 K (at 140 GPa) and 1915 K (at 211 GPa), respectively. By comparing the free energy difference, we determined the BCC→$RH_1$→$RH_2$→BCC boundaries and the BCC-$RH_1$-$RH_2$ triple point, which finally led us to construct a high-pressure and finite-temperature phase diagram for vanadium. The unexpected stabilization of the BCC phase by temperature was determined due to electronic entropy, which also leads to unusual hardening of the shear modulus $C_{44}$ by temperature, a very rare and interesting phenomenon.

By using the partial jellium model and intendedly adjusting the Fermi level position, we found that decreasing of the electronic chemical potential would further stabilizes the $RH_2$ phase at 211 GPa, whereas destabilizes it otherwise. By inspecting the changes in the density of states and band structures of vanadium when the Fermi level being shifted, we concluded that the most possible transition mechanism is a combination of the Fermi surface nesting, band splitting due to lattice distortions, and *d* orbitals localization. The former two affect the phase stability along the compression direction, while the latter favors BCC via the electronic entropy along the temperature dimension.

**Methods**

Most calculations are performed in the primitive cell of the BCC or RH phases, and the Vienna *ab initio* simulation package (VASP) is used,[29] which is based on first-principles density functional theory (DFT)[30] and the projector augmented-wave (PAW) method.[31] The pseudopotential contains 13 valence electrons (including $3s^2$, $3p^6$, $3d^3$, and $4s^2$ states). The Perdew-Burke-Ernzerhof (PBE) generalized gradient approximation (GGA)[32] for the electronic exchange-correlation functional is used. We speculate that the scattering in the previous theoretical results might relate to the calculation accuracy, therefore a high enough cutoff energy for the plane wave basis of 900 eV is used, as well as a $60\times60\times60$ shifted uniform mesh for the *k*-point sampling. This set includes the $\Gamma$ point and results in 5200 and 18910 *k*-points in the irreducible Brillouin zone of the BCC and RH lattices, respectively. The smearing parameters are also well tested. This parameter setting is carefully checked by increasing the cutoff energy and *k*-points to higher values and to ensure that it gives an absolutely converged total energy and pressure (in the sense of computer simulations).

In order to deliver a reliable energetics assessment, we exploit two different methods to evaluate the phase transitions and the (meta-)stability region. The first one is a conventional method (method I), in which we fully optimize the structure directly and then calculate the enthalpy of the resultant phases as a function of pressure. The transition pressures are obtained by comparing the enthalpy difference with respect to BCC phase. With this method, both the thermodynamic and mechanical (meta)stability are obtained. In the second method, we explore the structure transformation as that employed in Ref. 13, namely, distorting the BCC structure along a predefined pathway (method II). The instability and phase transitions are then deduced by inspecting the resultant energy curve as a function of the distortion magnitude. Different from the first one where structure is fully relaxed, in this method the explored geometry is highly constrained by the predefined pathway. Hence the absolute stability of these phases and their transition pressures are not guaranteed in principle. For this reason, Qiu *et al*. attributed the large discrepancy of their transition pressure from that of Lee *et al*. to the limitation of this method.[16] A cross-check of these two methods is necessary and will be helpful to secure a reliable and consistent theoretical result.

According to Ref. 13, the volume-conserved BCC→RH transformation matrix (or the deformation gradient) $T(\delta)$ is defined as

$$T(\delta) = \begin{pmatrix} k & \delta & \delta \\ \delta & k & \delta \\ \delta & \delta & k \end{pmatrix}, \qquad (2)$$

in which $k$ is determined from the real positive solution of det $(T) = 1$, to ensure a volume-conserving transformation. The small displacement $\delta$ represents the amount of rhombohedral deformation of the BCC crystal: a positive $\delta$ corresponds to a decrease in the angle $\alpha$ from the BCC value of $\alpha_0 = 109.47°$. This pathway, however, is not the one for the pure shear deformation. Therefore a different transformation matrix will be used when calculating $C_{44}$.

To evaluate the elastic moduli $C_{44}$ and $C' = (C_{11}-C_{12})/2$ of the BCC phase, we use a conventional 2-atom cubic unit cell to calculate the total energy as a function of volume and its variation along the shear strains. The $C_{44}$ and $C'$ are then obtained from the second derivative of the total energy with respect to the deformation magnitude $\delta$, which is defined by the strain matrices[33]

$$\varepsilon_{C_{44}} = \begin{pmatrix} 0 & \delta & 0 \\ \delta & 0 & 0 \\ 0 & 0 & \delta^2/(1-\delta^2) \end{pmatrix}, \tag{3}$$

$$\varepsilon_{C'} = \begin{pmatrix} \delta & 0 & 0 \\ 0 & -\delta & 0 \\ 0 & 0 & \delta^2/(1-\delta^2) \end{pmatrix}. \tag{4}$$

The corresponding strain energy is then given by

$$E(\delta) = E(0) + 2C_{44}V\delta^2 + O(\delta^4), \tag{5}$$

$$E(\delta) = E(0) + 2C'V\delta^2 + O(\delta^4). \tag{6}$$

As noted by Qiu et al.,[16] these formulae include the pressure corrections because of the $\delta^2$ term for $\varepsilon_3$, which they took as the main cause for the big deviation of their $C_{44}$ from that of Landa et al.[10]

At finite temperatures, one should use the free energy rather than the internal energy to derive the structure stability and physical properties. In general, the free energy as a function of temperature and density (or atomic volume) can be expressed as

$$F(V,T) = E_c(V) + F_e(V,T) + F_{vib}(V,T). \tag{7}$$

Here $E_c$ is the cold static lattice energy with atoms being clamped at their equilibrium positions, $F_e$ is the thermal free energy contributed by electronic excitations, and $F_{vib}$ is the vibrational free energy of phonons. Landa et al. calculated lattice dynamics and showed that $F_{vib}$ has little impact on the stability of RH phases.[21] For this reason we will mainly focus on the thermo-electrons effect in this work. After discarding the phonon term, the free energy becomes

$$F(V,T) = E_c(V) + F_e(V,T). \tag{8}$$

Within one-electron approximation, the thermo-electronic free energy $F_e(V,T)$ can be constructed from the groundstate density of states $n(\varepsilon)$ as

$$F_e(V,T) = E_e(V,T) - TS_e(V,T), \tag{9}$$

in which the internal energy due to electronic excitations is given by[34]

$$E_e(V,T) = \int_{-\infty}^{+\infty} \varepsilon n(\varepsilon, V) f(\varepsilon, T) d\varepsilon - \int_{-\infty}^{\varepsilon_F} \varepsilon n(\varepsilon, V) d\varepsilon, \quad (10)$$

and the electronic entropy is

$$S_e(V,T) = -k_B \int_{-\infty}^{+\infty} n(\varepsilon, V) \{f(\varepsilon, T) \ln f(\varepsilon, T) + [1 - f(\varepsilon, T)] \ln [1 - f(\varepsilon, T)]\} d\varepsilon, \quad (11)$$

where $f(\varepsilon,T)$ is the Fermi-Dirac distribution function, $n(\varepsilon,V)$ is the electronic density of states (DOS) at the energy eigenvalue $\varepsilon$, $\varepsilon_F$ is the Fermi energy, and $k_B$ is the Boltzmann's constant. This formalism can be devised and solved self-consistently with Mermin's finite temperature DFT.[25] For a given volume and temperature, from the variational principle of the free energy with respect to electron density, one first solves the standard Kohn-Sham equations[30] using a trial density. This produces one-electron eigenstates and the Fermi level. Then one reconstructs the charge density by populating electrons onto these states according to Fermi-Dirac distribution. Mixing this density with the initial one and recalculating the Kohn-Sham equations, then repeating the whole process until the convergence is achieved, one then obtains the self-consistent free energy of electrons at finite temperatures.[29] With this free energy, we can evaluate the thermo-electronic effect on phase stability and elastic constants.

Chemical environment or alloying will affect materials behavior by donating (or accepting) electrons. This phenomenon in vanadium has been explored in Landa *et al.*'s pilot works.[24,35] They found that the band-filling argument applies when alloying vanadium with its neighbors within the same transition series, whereas it is complicated by the Madelung energies in other cases. This precluded them from drawing a conclusion on the effects of charge transfer and electron chemical potential. In order to circumvent this difficulty and obtain an unambiguous understanding, we will study this problem with an alternative method, i.e., focusing on the effects solely due to the electron chemical potential. The electron chemical potential $\mu$, which is also the Fermi energy $\varepsilon_F$ in a metal, is defined as

$$\mu = \left(\frac{\partial F}{\partial N_e}\right)_{V,T}, \quad (12)$$

where $N_e$ is the total number of electrons. At the level of free electron gas model, its value at zero Kelvin is given by $\mu(0) = \hbar^2/2m_e \left(3\pi^2 N_e/V\right)^{2/3}$. Therefore one can adjust the Fermi-level position by adding or removing electrons from the system. In this work, we employ the partial jellium model to achieve this purpose by compensating the charged simulation cell with a homogeneous positive/negative background charge density. By allowing the added (fictitious) charges interacting with the nuclei and other electrons, the system relaxes and leads to a set of self-consistent eigenstates of the Kohn-Sham equations, from which the Fermi energy can be derived. With this method, the relaxation effect due to adding/removing electrons is automatically included. This method is widely used when studying charged defects in semiconductors.[26] Here we exploit it to investigate how the band-filling and Fermi-level position affect the high pressure behavior of vanadium. This will benefit us in understanding the phase transition mechanism, since different mechanisms usually have different response when the Fermi-level shifts.

**Acknowledgments**

This work is supported by the National Natural Science Foundation of China under Grant Nos. 11274281 and 11174214, the CAEP Research Projects under Grant Nos. 2102A0101001 and 2015B0101005, and the NSAF under Grant No. U1430117.

**Author Information**

Affiliations

National Key Laboratory of Shock Wave and Detonation Physics, Institute of Fluid Physics, CAEP; P.O. Box 919-102, Mianyang 621900, Sichuan, People's Republic of China.

Yi X. Wang, Q. Wu, and Hua Y. Geng

College of Physical Science and Technology, Sichuan University, Chengdu 610064, China.

Yi X. Wang and Xiang R. Chen

## Author Contributions

H.-Y.G. and Y.-X.W. conceived the research. Y.-X.W. performed atomic and electronic structure calculations. Y.-X.W., Q.W., X.-R.C., and H.-Y.G. analyzed the numerical results. Y.-X.W. and H.-Y.G. wrote the manuscript and all the authors commented on it.

## Competing Interests

The authors declare no competing financial interests.

## Corresponding Authors

Correspondence to Hua Y. Geng or Xiang R. Chen

Table I. Calculated critical pressures of vanadium at zero Kelvin, in which $P_i$ denotes the lower bound of the metastable pressure, $P_c$ is the thermodynamically equilibrium transition pressure, and $P_m$ is the pressure where $RH_2$ phase attains the greatest stability, respectively.

| Critical pressure | Method I (GPa) | Method II (GPa) | Lee et al.[13] (GPa) | Qiu et al.[16] (GPa) | Verma et al.[15] (GPa) |
|---|---|---|---|---|---|
| $P_i$ of $RH_1$ | 50 | — | 73 | 19 | — |
| $P_c$ of BCC→$RH_1$ | 98 | 103 | 84 | 32 | 60 |
| $P_i$ of $RH_2$ | 110 | — | 103-112 | 65 | — |
| $P_c$ of $RH_1$→$RH_2$ | 128 | 126 | 119 | 115 | 160 |
| $P_m$ of $RH_2$ | 210 | 211 | 187 | — | — |
| $P_c$ of $RH_2$→BCC | 284 | 278 | 280 | 297 | > 240 |

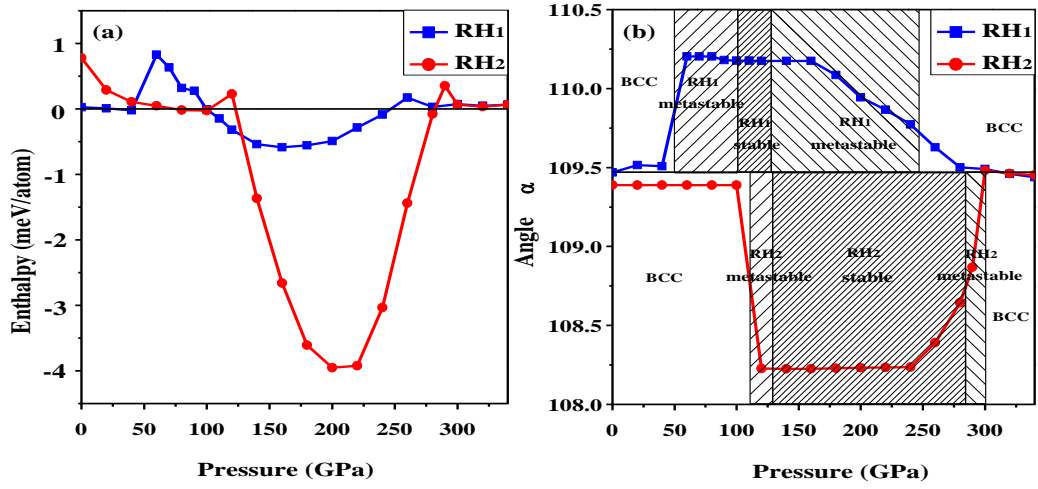

Figure 1. (Color online) (a) Enthalpy difference of vanadium in $RH_1$ and $RH_2$ structures at zero Kelvin with respect to the BCC phase as a function of pressure. (b) Variation of angle $\alpha$ in $RH_1$ and $RH_2$ structures as a function of pressure at zero Kelvin. Note that $\alpha = 109.47°$ corresponds to the perfect BCC structure.

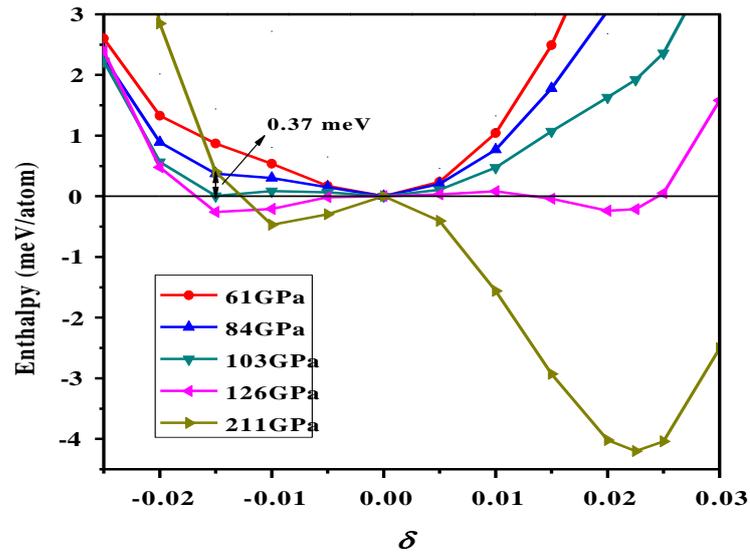

Figure 2. (Color online) Variation of the calculated enthalpy difference as a function of the rhombohedral deformation parameter $\delta$ at selected pressures. The negative $\delta$ on the left side corresponds to the RH$_1$ phase, and the positive $\delta$ on the right side is for the RH$_2$ phase.

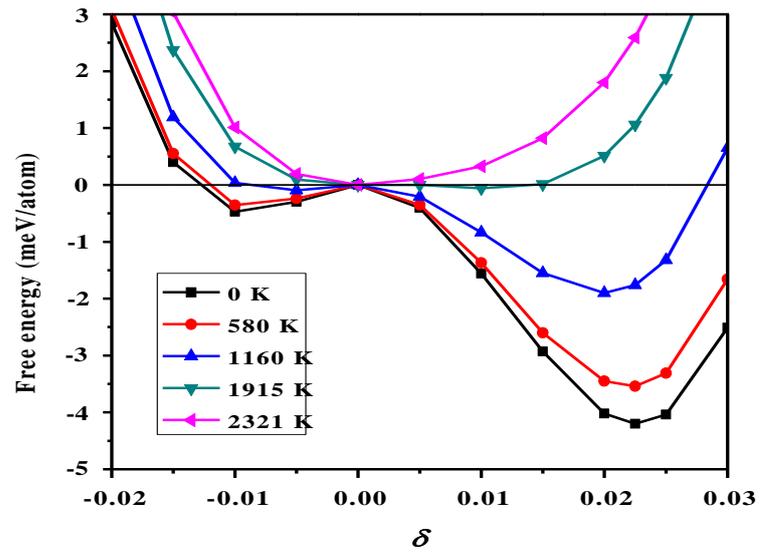

Figure 3. (Color online) Effect of thermo-electrons on the phase stability of vanadium in the $RH_1$, $RH_2$, and BCC structures at 211 GPa.

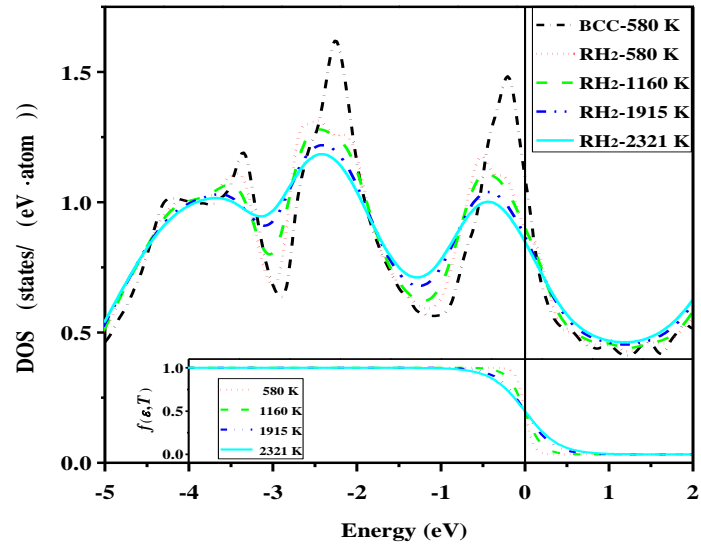

Figure 4. (Color online) Comparison of the electronic density of states of vanadium at 211 GPa for BCC phase at 580 K and RH$_2$ phase at various temperatures. The inset shows the Fermi-Dirac distribution function at the given temperatures. The vertical line denotes the Fermi level.

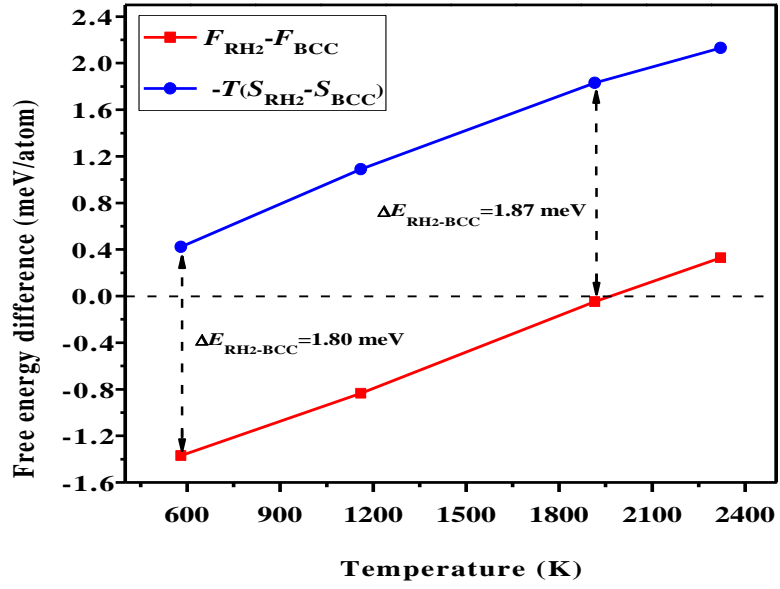

Figure 5. (Color online) Variation of the free energy difference and its entropy contribution (-*TS*) between RH$_2$ (with $\delta$ = 0.01) and BCC phase of vanadium as a function of temperature at a pressure of 211 GPa. The dashed lines with arrowheads denote the difference of internal energy.

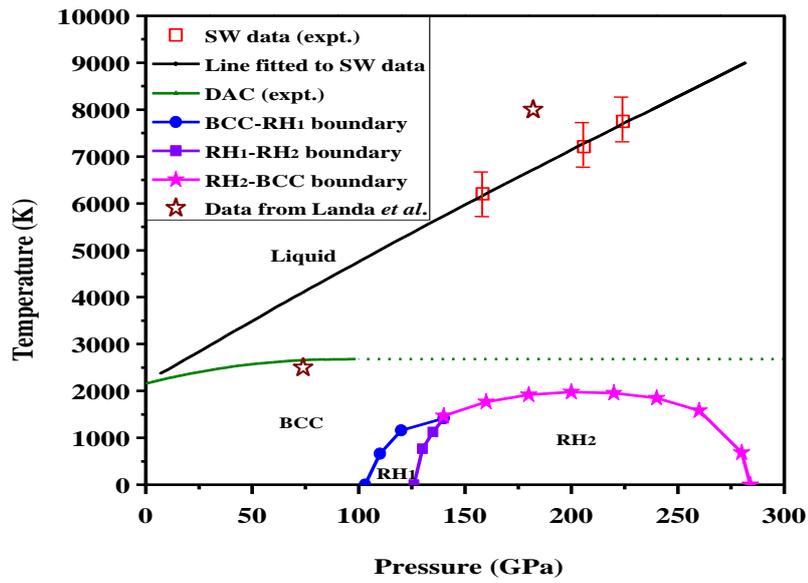

Figure 6. (Color online) Phase diagram of vanadium at high pressure and finite temperature. The DAC data of the melting line and its extrapolation are taken from Ref. [22]. The shock wave (SW) data on the melting curve are from Ref. [23]. The stability bounds of RH phases estimated by Landa *et al.* are from Ref. [21].

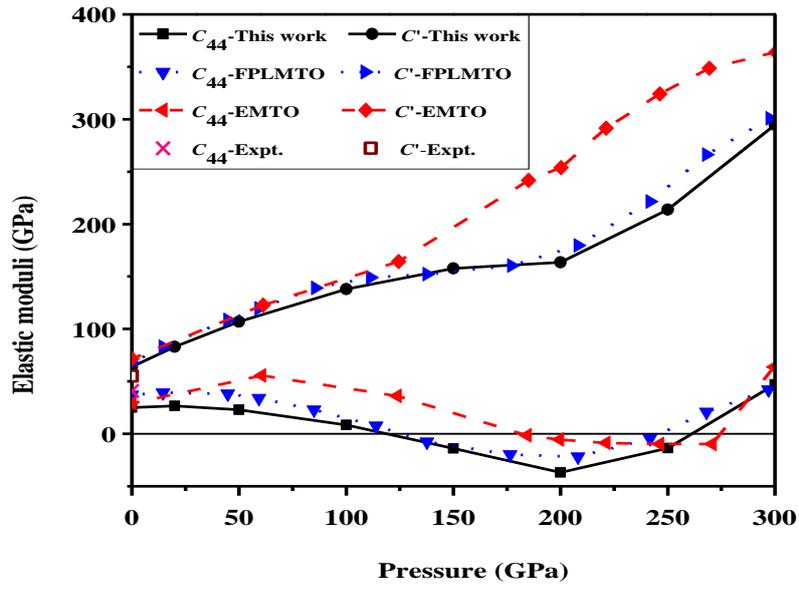

Figure 7. (Color online) Elastic moduli of vanadium in BCC phase as a function of pressure. The results of $C_{44}$-FPLMTO, $C'$-FPLMTO, $C_{44}$-EMTO, and $C'$-EMTO are taken from Ref. [10]. The experimental data at ambient conditions are from Ref. [26].

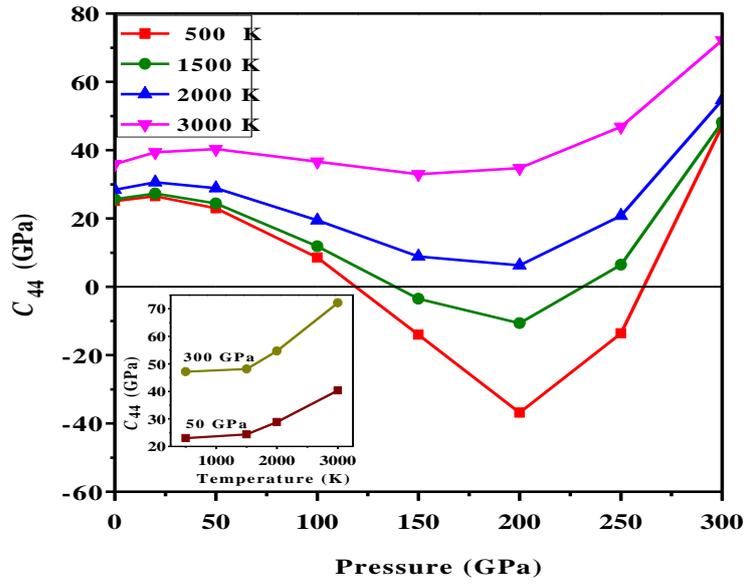

Figure 8. (Color online) Calculated shear elastic constant $C_{44}$ of vanadium in BCC phase as a function of pressure at different electronic temperatures. Inset: Variation of $C_{44}$ as a function of temperature at given pressures of 50 and 300 GPa, respectively.

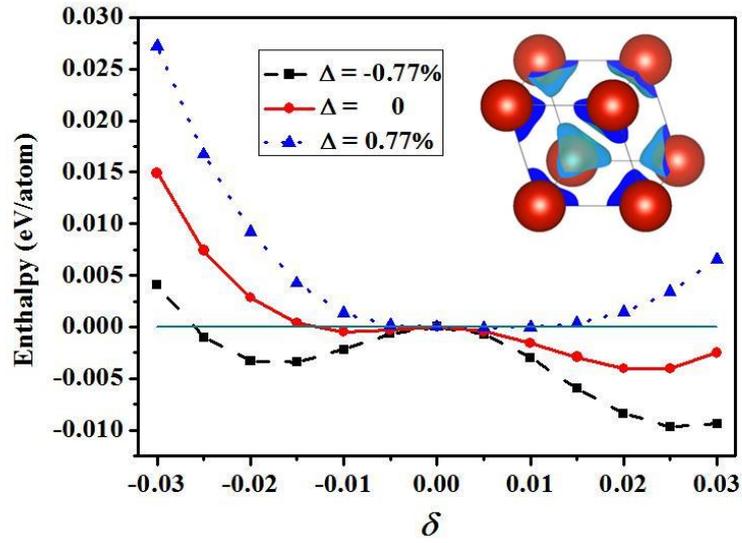

Figure 9. (Color online) Variation of the enthalpy difference with respect to BCC phase as a function of the rhombohedral deformation parameter $\delta$ when the Fermi level being shifted up or down at a pressure of 211 GPa. Inset: Calculated differential charge density between $\Delta$ = -0.77% and $\Delta$ = 0. Here $\Delta$ is the percentage of the total charge that are removed from (or added to) the system for the purpose to shift the Fermi level.

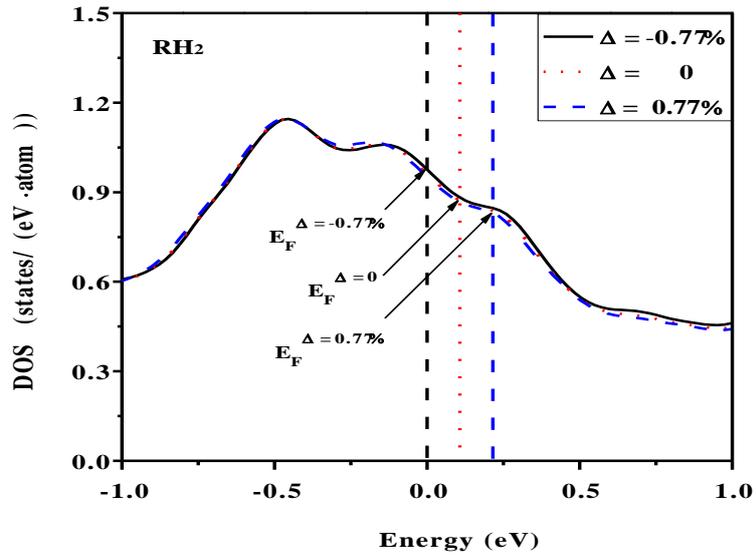

Figure 10. (Color online) Variation of the electronic density of states and the position of the Fermi level when additional electrons or holes are added to vanadium in RH$_2$ phase at 211GPa.

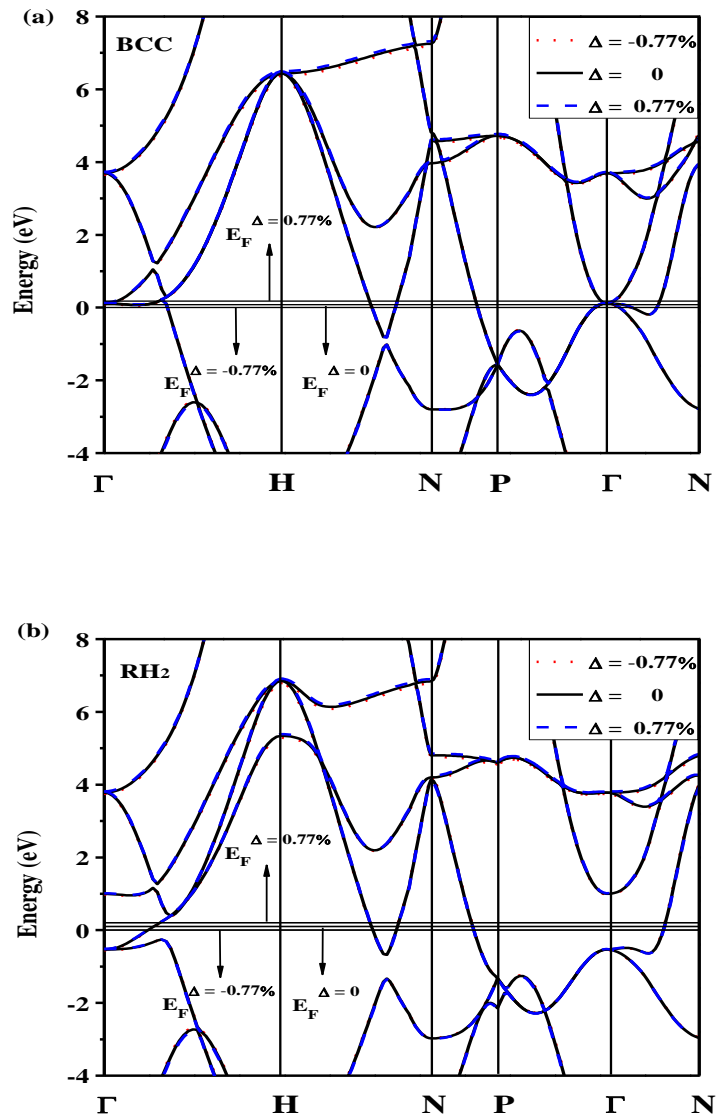

Figure 11. (Color online) Band structures of vanadium at 211GPa when the Fermi level is shifted: (a) BCC phase, (b) RH$_2$ phase.

Supplementary information for

# Stability of rhombohedral phases in vanadium at high-pressure and high-temperature: first-principles investigations

Yi X. Wang,[1,2] Q. Wu,[1] Xiang R. Chen,[2,*] and Hua Y. Geng[1,*]

[1]*National Key Laboratory of Shock Wave and Detonation Physics, Institute of Fluid Physics, CAEP; P.O. Box 919-102, Mianyang 621900, Sichuan, People's Republic of China.*
[2]*College of Physical Science and Technology, Sichuan University, Chengdu 610064, China.*
*e-mail: s102genghy@caep.ac.cn; xrchen@scu.edu.cn

**Lattice vibrations correction on the phase stability of vanadium.** By using the finite temperature DFT method, we have investigated the thermo-electronic effects on the phase stability of vanadium in this work. It is an unusual approach to include electronic excitations but ignore the phonon contributions. We are indeed aware of the importance of lattice vibrations on phase stability of solid. But we also realized by empirical experiences that phonons' contribution is not large for some metals, and its impact on the phase diagram can be included via an approach similar to perturbation correction. In the case of vanadium we considered here, we knew that both phonon and thermo-electronic effects could affect the RH stability. The basis of our approach to ignore lattice vibrations actually relys on Landa's phonon calculation (Ref. 21 of the main text), which demonstrated that phonons' impact on RH stability is not big. We took this as an assumption and calculated the thermo-electronic effects separately. There are two cases one could expect: (i) the thermo-electronic effects are at the comparable level with the lattice vibrations, thus invalidates our assumption that phonons contribution is negligible, and a full consistent treatment containing both lattice vibrations and electronic excitations is required; and (ii) the thermo-electronic effects are much more stronger, and validates our assumption that phonons contribution can be treated as a perturbation. Comparing our thermo-electronic results and Landa's lattice dynamics confirms that the case (ii) is true. For many metals, thermo-electronic effects are usually smaller than phonons contribution. But our discovery here is that for vanadium, it is opposite. In order to eliminate the worry about the validation of our conclusion, we present a simple assessment of the phonons contribution below.

Starting from the definition of the Helmholtz free energy of a solid

$$F(V,T) = E_c(V) + F_e(V,T) + F_{vib}(V,T). \tag{1}$$

in which $E_c$ stands for cold energy, $F_e$ stands for the free energy from thermal electrons and $F_{vib}$ stands for the lattice dynamics contribution, respectively. The phase diagram given in Ref. 21 implies that at the given phase boundary between BCC and RH phase at 182 GPa and 8000 K, there is

$$E_{c,RH} - E_{c,BCC} = \Delta E_c = F_{vib,BCC} - F_{vib,RH} = -\Delta F_{vib}(T = 8000\,\text{K}). \tag{2}$$

On the other hand, if including the thermo-electronic effects only, our finite temperature phase diagram implies

$$\Delta E_c = F_{e,BCC} - F_{e,RH} = -\Delta F_e(T \approx 2000\,\text{K}). \tag{3}$$

Since the Debye temperature of vanadium is about 326 K at the ambient conditions. Compression to high pressures about 200 GPa increases the Debye temperature to a level of 600 K, but is still far smaller than the temperature scale we are considering here. This implies that we can use the high temperature expansion to approximate the lattice vibrational free energy $F_{vib}$. Namely

$$F_{vib}(T) = -3k_B T \left[ \ln\left(\frac{T}{\theta_0}\right) - \frac{1}{40}\left(\frac{\theta_2}{T}\right)^2 + \cdots \right] \approx -3k_B T \ln\left(\frac{T}{\theta_0}\right). \tag{4}$$

Therefore

$$-F_{vib}(T) = -3k_B T \ln\left(\frac{\theta_{0,RH}}{\theta_{0,BCC}}\right). \tag{5}$$

and

$$\frac{F_{vib}(T = 8000\text{K})}{F_{vib}(T = 2000\text{K})} = 4. \tag{6}$$

This gives an estimation of the lattice free energy difference at 2000 K, which is about one fourth of that at 8000 K, and thus about one fourth of $\Delta F_e(T = 2000\,\text{K})$. It is evident that neglect of the lattice vibrational contribution overestimates the transition temperature. An approach to correct this and include the phonon effect on the phase boundary is to consider the free energy of thermal electrons as a function of temperature. For most metals, the free electron gas model can be applied and the electronic free energy is

$$F_e = -\frac{1}{2}\Gamma T^2, \tag{7}$$

in which $\Gamma$ is a constant. Thus $\Delta F_e \propto T^2$. At the real transition temperature $T'$, one has

$$\Delta F(T') = \Delta E_c + \Delta F_e(T') + \Delta F_{vib}(T') = 0. \tag{8}$$

Approximate $\Delta F_{vib}(T')$ by $\Delta F_{vib}(T = 2000\,\text{K})$, one gets

$$-\Delta F_e(T') \approx -\frac{3}{4}\Delta F_e(T = 2000\,\text{K}). \tag{9}$$

Eq. (7) then gives

$$T' \approx \sqrt{\frac{3}{4}} \times 2000\,\text{K} = 1732\,\text{K}, \tag{10}$$

or

$$\frac{T'-T}{T} = \frac{\Delta T}{T} = 0.13. \tag{11}$$

It shows that if include only the thermal electrons and neglect lattice vibrations, the phase transition temperature will be overestimated by about 13% (or ~268 K at ~200 GPa). This confirms the argument that the stability of RH will be reduced further by phonon effects. But the correction magnitude is actually small. A version of the corrected phase diagram is given in the Fig. S1. The concern that including both thermo-electron effects and phonon contribution might greatly alter the phase diagram should not exist.

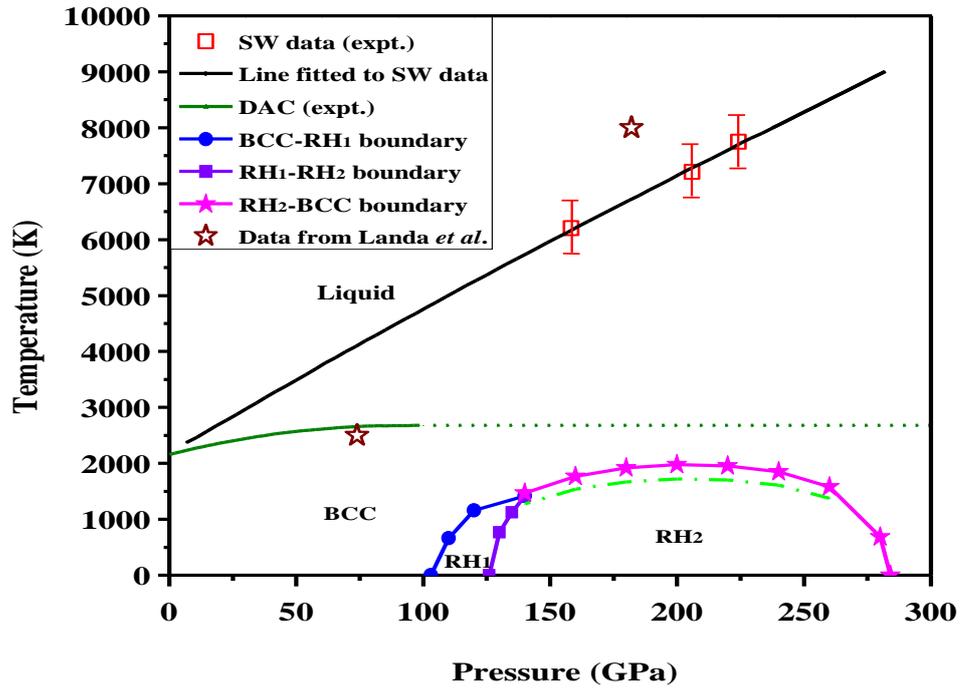

Figure S1. (Color online) Phase diagram of vanadium at high pressure and finite temperature as shown in the Fig. 6 of the main text. The dash-dotted line indicates the correction of lattice dynamics if included via a simple assessment.